\documentclass[twocolumn]{aastex631}

\usepackage{amsmath}
\usepackage{bm}
\usepackage{bbding}

\shorttitle{}
\shortauthors{Xia et al.}

\begin{document}

\title{First Evidence for a QPO Triplet and Its Relativistic Precession Origin in RE J1034+396}
\author[0009-0005-3916-1455]{Ruisong Xia}
\author[0000-0002-1935-8104]{Yongquan Xue\textsuperscript{\Envelope}}
\author[0009-0001-9170-3363]{Jialai Wang\textsuperscript{\Envelope}}
\author[0000-0001-5525-0400]{Hao Liu\textsuperscript{\Envelope}}

\affiliation{Department of Astronomy, University of Science and Technology of China, Hefei 230026, China; xuey@ustc.edu.cn, jialaiwang@mail.ustc.edu.cn, liuhao1993@ustc.edu.cn}
\affiliation{School of Astronomy and Space Science, University of Science and Technology of China, Hefei 230026, China}

\begin{abstract}
Quasiperiodic oscillations (QPOs) in active galactic nuclei (AGNs) provide a powerful tool for probing the structure of the innermost accretion flow and corona around supermassive black holes.
RE~J1034+396, the most prominent AGN known to host an X-ray QPO, exhibits both short-term and long-term QPO evolution, offering a unique opportunity to investigate accretion disk and corona physics through its temporal behavior.
We report a possible long-term ($\sim 92.2$ days) cyclic evolution of the QPO in RE~J1034+396, joining the detected QPO ($\sim 3730$ s) and its short-term ($\sim 17$ ks) modulation to form a possible QPO triplet, which is potentially the first such structure identified in an AGN.
By applying the relativistic precession model {to the QPO triplet}, we constrain the black hole mass to $1.7^{+0.9}_{-0.8} \times 10^{6}\ M_\odot$, consistent with independent estimates, and find a low dimensionless {black hole} spin of $0.017^{+0.028}_{-0.012}$.
We propose an exploratory model {that involves a quasiperiodic ultra-fast outflow (UFO) within the framework of the relativistic precession model}, explaining the QPO lag reversal, the modulation of hard-band QPO amplitude by soft-band flux, and the long-term evolution of timing properties.
Supporting evidence includes blueshifted emission and absorption lines indicating a strong UFO at $\sim 0.3c$.
This work provides new insights into the inner regions of AGN accretion disks and motivates further efforts in both numerical modeling and high-cadence timing observations.

\end{abstract}


\section{Introduction} \label{sec:intro}

When gas is accreted into the inner region of the accretion disk {that lies at the immediate vicinity of} the black hole, it may be evaporated into a corona \citep{1999ApJ...527L..17L} before being drawn into the black hole.
As a bulk of high-temperature plasma, the corona emits strong X-rays.
It dominates the X-ray emission of {active galactic nuclei} (AGNs).
The structure of the corona and its interplay with the accretion process have been longstanding subjects of investigation.
While most X-ray observations of AGNs exhibit stochastic variability, a small subset shows X-ray quasi-periodic oscillations (QPOs), indicating the presence of coherent processes that may reflect an underlying pattern in the evolution of the coronal structure.

The structure and variability of the corona region in RE~J1034+396, a narrow-line Seyfert 1 galaxy (NLS1) of our focus, could be distinctive, characterized by the most pronounced X-ray QPO among AGNs \citep{2008Natur.455..369G} and its {unique evolutionary} properties \citep{2020MNRAS.495.3538J, 2024ApJ...961L..32X, 2025ApJ...983...13X}.
Since the first discovery of a QPO in RE~J1034+396, which was also the first such detection among AGNs \citep{2008Natur.455..369G}, numerous hypotheses have been proposed to explain this remarkable phenomenon.
{For example, using} phase-resolved spectroscopy, \citet{2010ApJ...718..551M} identified evidence for a warm absorber that periodically obscures the continuum emission.  
\citet{2011MNRAS.414..627D} developed a model in which the QPO arises from oscillations of a shock formed in a low-angular-momentum, hot accretion flow.  
\citet{2011MNRAS.417..250M} and \citet{2021MNRAS.500.2475J} proposed a physical analogy between RE~J1034+396 and the black hole binary (BHB) GRS 1915+105, suggesting that similar mechanisms may drive the observed QPO {behaviors} across different black hole mass scales.

BHBs, often referred to as {microquasars}, possess accretion systems similar to those of AGNs but are centered on stellar-mass black holes.
QPOs have been studied much more extensively in BHBs, and various models have been proposed to explain their origin \citep{2019NewAR..8501524I}.
The relativistic precession model (RPM) \citep{1998ApJ...492L..59S, 1999ApJ...524L..63S} could be the simplest {among} all the models.
It predicts QPO triplets {that are associated with the Lense-Thirring (LT)} precession \citep{1918PhyZ...19..156L}, periastron precession, and orbital motion, all at the same radius \citep{2014MNRAS.437.2554M, 2019NewAR..8501524I}.

The QPO in RE~J1034+396 is not consistently stable. 
A drift in the QPO central frequency was observed \citep{2010A&A...524A..26C}, correlating with delayed changes in the X-ray flux.
Additionally, a long-term variation was studied by \citet{2020MNRAS.495.3538J}.
\citet{2024ApJ...961L..32X} reported a hysteretic evolution between the QPO lag $\tau_\mathrm{QPO}$ and frequency $f_\mathrm{QPO}$, suggesting a potential long-term evolutionary cycle.
\citet{2025ApJ...983...13X} further organized {and analyzed its 12 long-exposure} observations according to their phases within this hypothesized cycle.  
Additionally, \citet{2025ApJ...983...13X} identified a lower-frequency QPO at $\rm (5.25\pm 1.86) \times 10^{-5}\ Hz$, corresponding to the $\sim 17\ \rm ks$ QPO modulation.  
These long-term ($\sim 92.2\ \rm days$; estimated in Section~\ref{Pdiag}) and short-term ($\sim 17\ \rm ks$) QPO behaviors, along with the QPO itself ($\sim 3730\ \rm s$), may all be part of a QPO triplet.

In this paper, we propose an exploratory model based on the RPM that simultaneously accounts for the long-term evolution \citep{2024ApJ...961L..32X}, the short-term evolution \citep{2025ApJ...983...13X}, and the QPO itself \citep{2008Natur.455..369G}.
In Section~\ref{sec_model}, we apply the RPM to the possible QPO triplet to infer the black hole mass and spin of RE~J1034+396, and outline a corresponding scenario for the observed evolutionary properties.
We search for supporting evidence for this hypothesis by analyzing spectral features in Section~\ref{sec_spec}.
The model and findings are discussed in Section~\ref{sec_dis} and summarized in Section~\ref{sec_sum}. {The analysis in this work is based on the same dataset as \citet{2025ApJ...983...13X}, i.e., the 10 observations (Obs-1 to Obs-10) taken between 20 November 2020 and 30 May 2021 as well as Obs-a taken in 31 May 2007 and 0bs-b taken 30 October 2018.}

\section{The Model}\label{sec_model}
\begin{figure}
    \centering
    \includegraphics[scale=0.4]{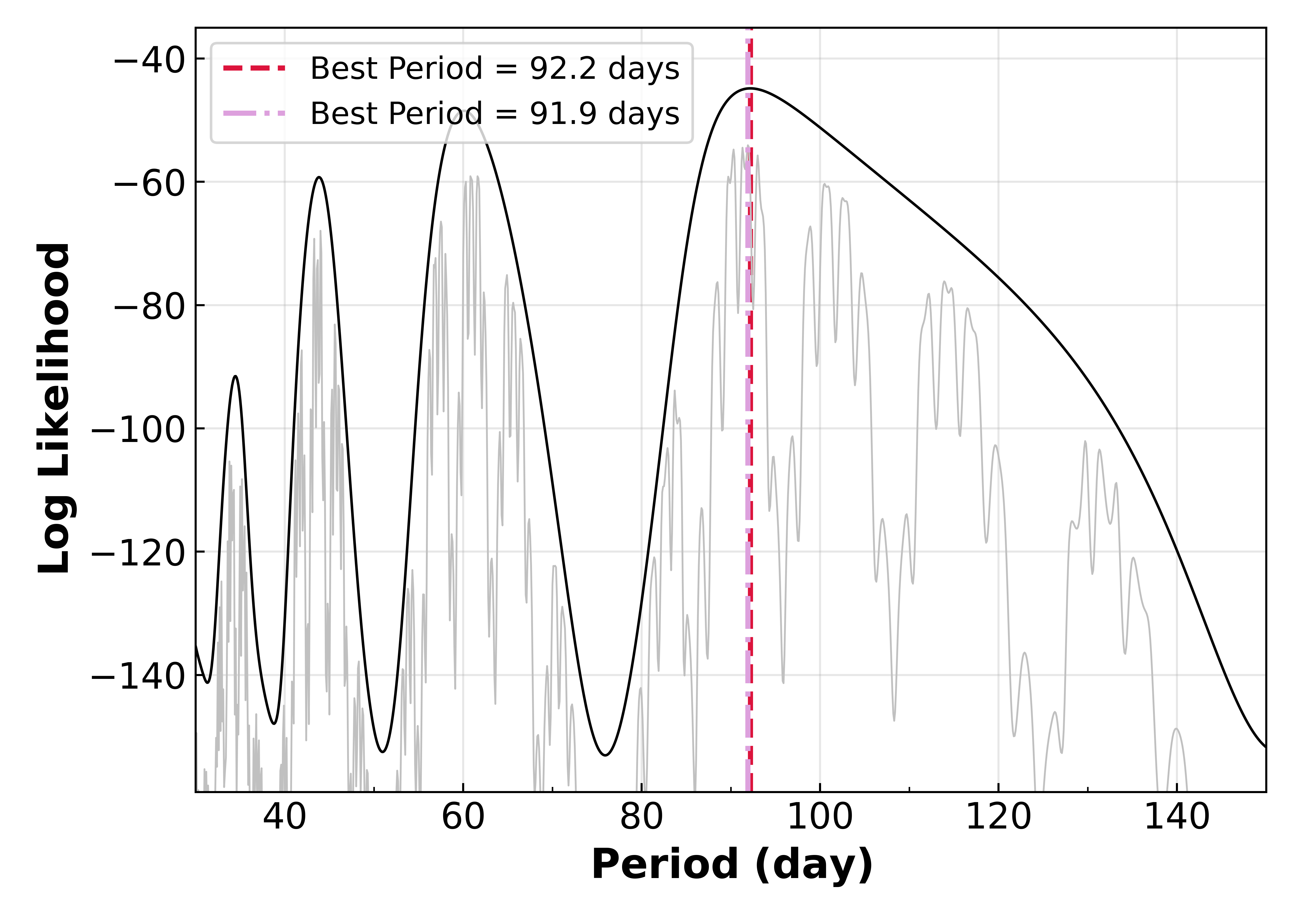}
    \caption{{Log-likelihood as a function of trial period, shown by the black (gray) curve for the 10 (12) observations. The red (pink) dashed line indicates the corresponding maximum.}
    }
    \label{fig1}
\end{figure}

\begin{figure*}
    \centering
    \includegraphics[scale=0.6]{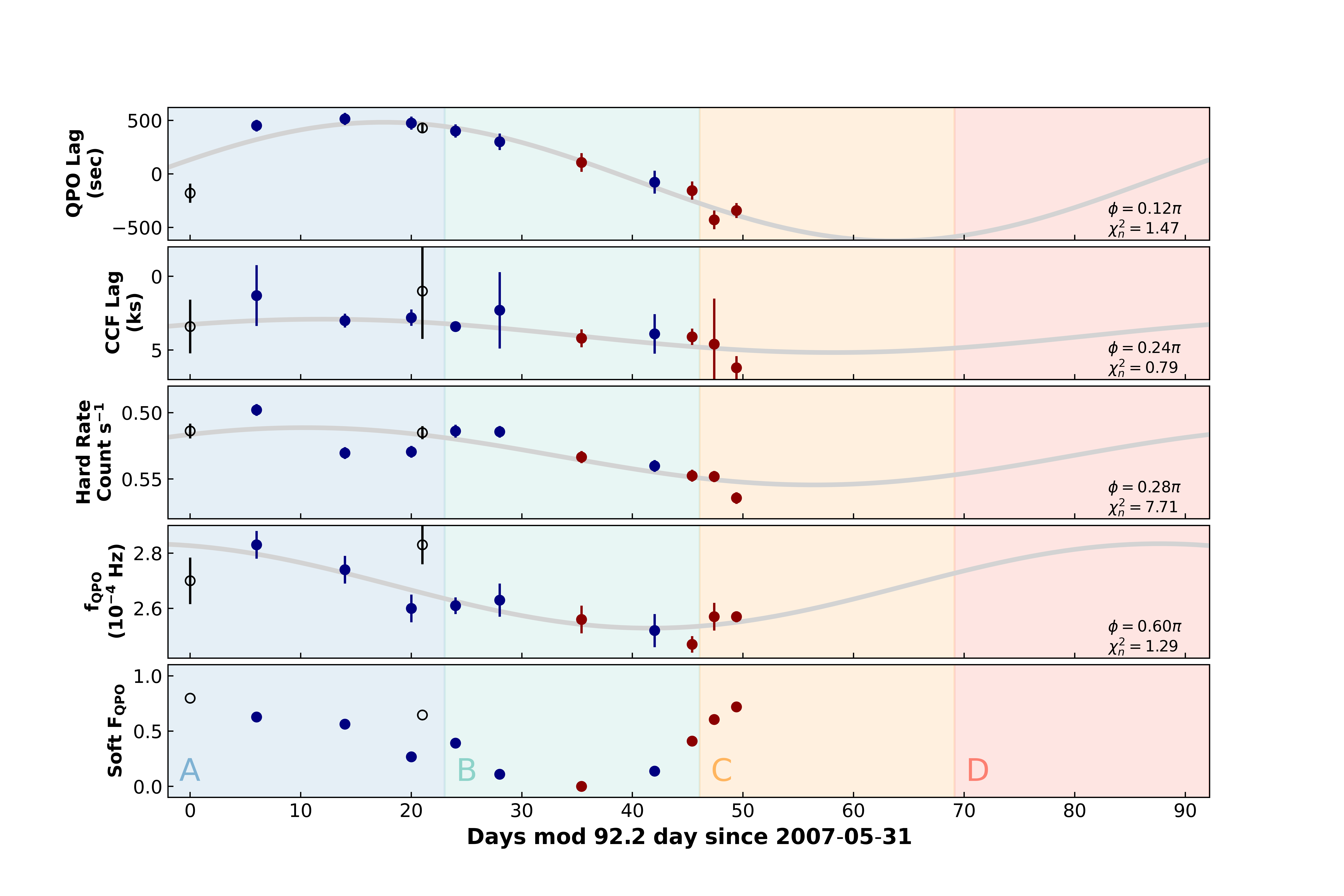}
    \caption{The period-folded evolution of the QPO lag, CCF lag (i.e., the lag between the smoothed soft flux and the hard QPO amplitude; \citealt{2025ApJ...983...13X}), hard-band count rate, QPO frequency ($f_{\rm QPO}$), and the detection fraction of the soft QPO (soft $F_{\rm QPO}$).
    {Dark red points represent data from November to December 2020 (i.e., Obs-1 to Obs-4), while dark blue points represent data from April to May 2021 (i.e., Obs-5 to Obs-10). The two earlier observations (Obs-a and Obs-b) are shown as open black circles for reference and are not included in the fitting.}
    A sinusoidal model is fitted to each parameter, except for the soft $F_{\rm QPO}$.  
    The grey curves indicate the best-fit models, and the corresponding phase and reduced chi-squared values are labeled in the lower right corner of each panel.
    {We artificially divide the evolution into four intervals to represent the typical states of the parameters, labeled as colored regions A, B, C, and D.}
    }
    \label{fig2}
\end{figure*}

\begin{figure}
    \centering
    \includegraphics[scale=0.5]{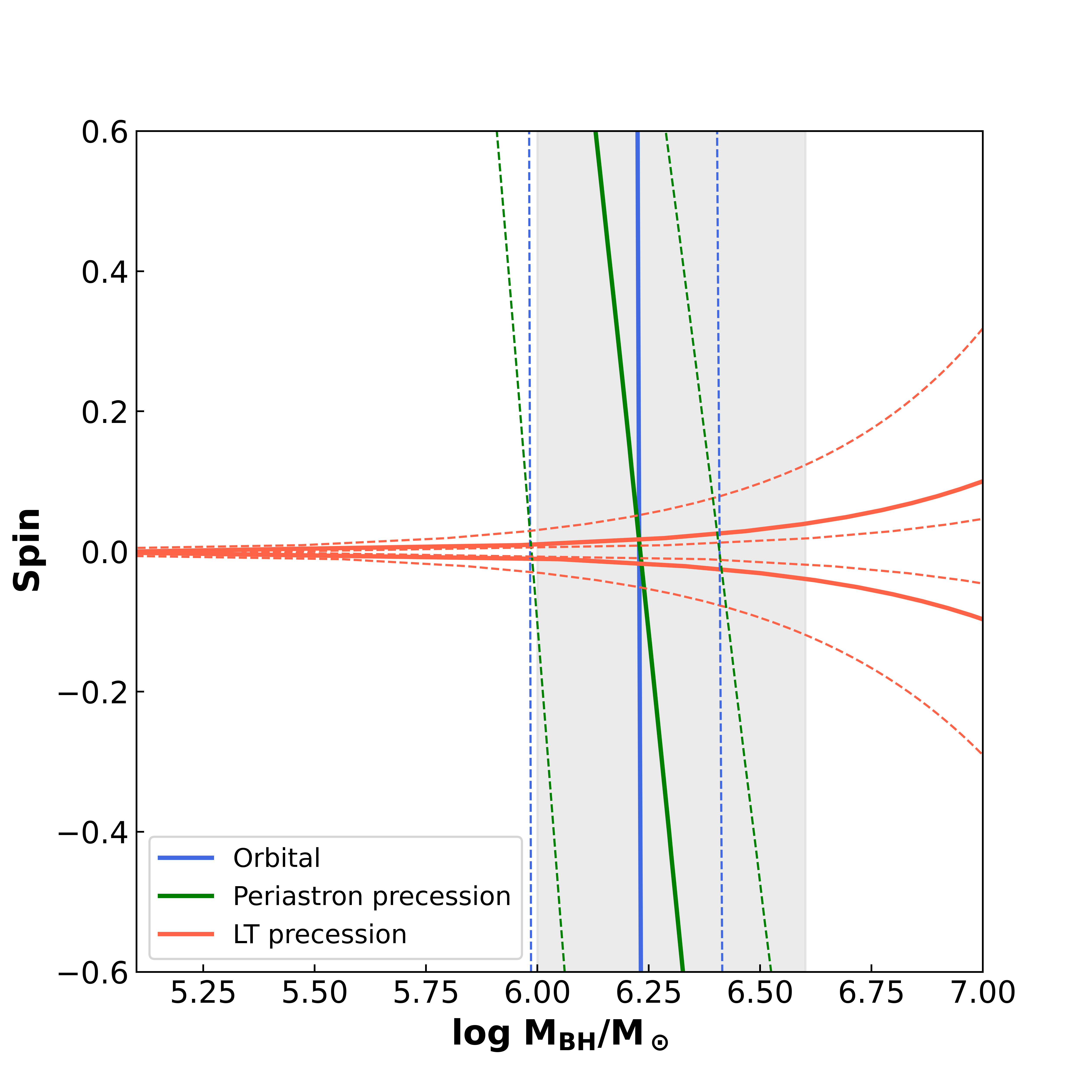}
    \caption{Black hole spin as a function of {black hole} mass at a radius of $r=17.1^{+3.8}_{-7.8}$, as predicted by the three equations of the RPM, in a diagram analogous to \citet{2014MNRAS.437.2554M}, but applied to RE~J1034+396.  
    The blue, green, and red curves represent the orbital, periastron precession, and {LT} {precession} frequencies, respectively.  
    Solid {curves} indicate the best-fit values, while dashed {curves} show the 1$\sigma$ confidence {intervals} for the mass measurement.
    {We derive a black hole mass of $1.7^{+0.9}_{-0.8} \times 10^{6}\ M_\odot$ and a black hole spin of $0.017^{+0.028}_{-0.012}$. 
    The derived mass is consistent with the independent estimate based on stellar velocity dispersion \citep{2009ASPC..408..303B, 2010MNRAS.401..507B} within the $1\sigma$ range (gray area).}
    }
    \label{fig3}
\end{figure}

\subsection{Periodicity Diagnosis} \label{Pdiag}

In the RPM, the origin of the QPO is attributed to orbital motion, {with the orbital frequency given by \citet{2019NewAR..8501524I} as}
\begin{equation} \label{eq-phi}
\nu_\phi = \pm \frac{1}{2\pi}\frac{c}{R_g}\frac{1}{r^{3/2}\pm a},
\end{equation}
{where $c$ is the speed of light, $a$ is the dimensionless black hole spin parameter}, and $r$ is the radial coordinate in units of the gravitational radius, i.e., $r = R / R_g$.
Due to relativistic effects in the Kerr metric, the radial and vertical epicyclic frequencies deviate from the orbital frequency \citep{2019NewAR..8501524I}:
\begin{equation}
\nu_r = \nu_\phi \sqrt{1 - \frac{6}{r} \pm \frac{8a}{r^{3/2}} - \frac{3a^2}{r^2}},
\end{equation}
\begin{equation}
\nu_z = \nu_\phi \sqrt{1 \mp \frac{4a}{r^{3/2}} + \frac{3a^2}{r^2}}.
\end{equation}
These deviations give rise to periastron precession and {the LT} precession \citep{2019NewAR..8501524I}, defined as:
\begin{equation} \label{eq-per}
     \nu_\mathrm{per} = \nu_\phi - \nu_r,
\end{equation}
\begin{equation} \label{eq-LT}
    \nu_\mathrm{LT} = \nu_\phi - \nu_z,
\end{equation}
which may correspond to the short- and long-term evolution of the QPO, respectively.

To search for possible periodicity under the assumption of a {long-term cyclic} evolution \citep{2024ApJ...961L..32X, 2025ApJ...983...13X}, we perform a sinusoidal fit of the form
\begin{equation}
 m(t) = A \sin\left( \frac{2\pi t}{P} + \phi \right),
\end{equation}
to the {QPO lags between the 0.3--1 and 1--4 keV bands (soft and hard bands, respectively) of the 10 observations previously measured by} \citet{2024ApJ...961L..32X} over a grid of trial periods from 50 to 180 days with a resolution of 0.1 days.
Here, $A$ is the amplitude, $\phi$ is the phase offset, and $P$ is the trial period.
The model is fitted directly to each observed parameter $y_i$ at observation times $t_i$ by maximizing the Gaussian log-likelihood:
\begin{equation}
\ln \mathcal{L} = -\frac{1}{2} \sum_i \left[ \frac{(y_i - m(t_i))^2}{\Delta y_i^2} + \ln(2\pi \Delta y_i^2) \right],
\end{equation}
where $m(t_i)$ is the model prediction for the parameter at time $t_i$, and $\Delta y_i$ is the corresponding measurement uncertainty.
For each trial period, the amplitude and phase are optimized using the \texttt{scipy.optimize.minimize} function to maximize the likelihood.
The resulting log-likelihood values are recorded as a function of trial period ({see} Figure~\ref{fig1}), with the highest likelihood occurring at 92.2 days ({corresponding to} $1.26\times10^{-7}\ \rm Hz$), which is adopted as the candidate best-fit period.
Figure~\ref{fig2} presents the parameters folded with this period and the corresponding best-fit sinusoidal curves, highlighting clear evolutionary trends, particularly in the QPO lag.
{Using 1000-step block bootstrap resampling with a block size of 3, we estimate the uncertainty in the period to be approximately 5.2 days.
}
Note that performing a joint fit to the QPO lag, CCF lag (i.e., the lag between the smoothed soft{-band} flux and the hard{-band} QPO amplitude estimated via the cross-correlation function, hereafter referring specifically to this definition; \citealt{2025ApJ...983...13X}), hard-band count rate, and QPO frequency \citep{2024ApJ...961L..32X, 2025ApJ...983...13X}, by maximizing the sum of their individual log-likelihoods, yields a best-fit period {that differs from 92.2 days by less than 0.1 day.}
Additionally, we validate that including the two earlier observations (Obs-a and Obs-b) does not significantly affect the fitted period as well {(see Figure~\ref{fig1})}.
{Therefore, we adopt a hypothetical period of $92.2 \pm 1.6$ days for the long-term evolution.}

\begin{figure*}[p]
    \centering
    \includegraphics{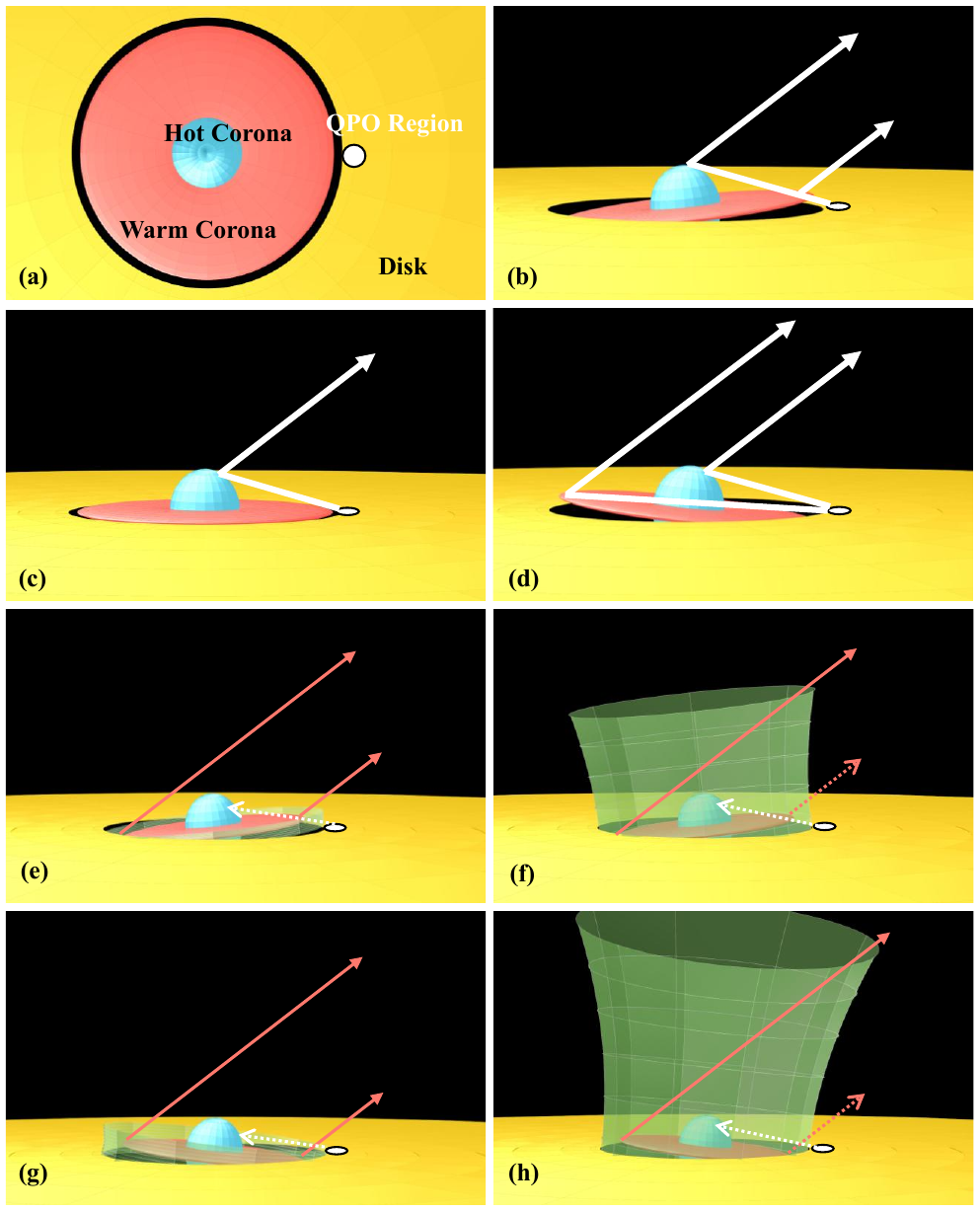}
    \caption{Schematic illustrations of the QPO evolution model {(plotted using \citealt{The_Manim_Community_Developers_Manim_Mathematical_2025})}.
    (a) Hypothetical structure of the corona (top-down view): the warm corona (red), the hot corona (blue), and the seed-photon-emitting region (white, {with black circle as boundary}) on the accretion disk (yellow) associated with the QPO.  
    (b)–(d) Long-term evolution of the system (front view). As a result of {LT} precession, the tilt angle between the coronae and the disk evolves to $-\beta$ (b), $0$ (c), and $\beta$ (d). The trajectories of the QPO-associated photons are shown by white arrows.  
    (e)–(h) Short-term evolution (front view) when the tilt angle is $-\beta$ (e and f) and $\beta$ (g and h). 
    The green spheres represent structures formed by clumpy outflows.
    Red arrows denote soft X-ray photons, while dotted lines indicate photon paths that are obscured by the outflow. 
    The observer is located at the top-{right} corner of the schematics (b-h).
    }
    \label{fig4}
\end{figure*}
By simultaneously solving Equations~\eqref{eq-phi},~\eqref{eq-per}, and~\eqref{eq-LT} with {observational inputs of} $\nu_\phi=2.65\times10^{-4}\ \rm Hz$ (QPO frequency), $\nu_\mathrm{per}=5.25\times10^{-5}\ \rm Hz$ (the short-term evolution period inferred from stacked power spectral densities, associated with the intermittent QPO timescale; see Section~3.2 and Appendix~A of \citealt{2025ApJ...983...13X}), and $\nu_\mathrm{LT}=1.26\times10^{-7}\ \rm Hz$, we derive a black hole mass of $M_\mathrm{BH}=1.7^{+0.9}_{-0.8} \times 10^6 M_\odot$, a dimensionless radius of $r=17.1^{+3.8}_{-7.8}$, and a dimensionless spin parameter of $a=0.017^{+0.028}_{-0.012}$.
Following the approach of \citet{2014MNRAS.437.2554M}, the uncertainties are estimated using the Monte Carlo method, based on 1000 sets of the three frequencies resampled according to their measured uncertainties.
Figure~\ref{fig3} shows the black hole spin as a function of {black hole} mass, as predicted by the three equations of the RPM at the {radius of $r=17.1^{+3.8}_{-7.8}$}.
The physically reasonable radius, along with the consistency between the previously measured black hole mass and the results of the periodicity analysis, supports interpreting the QPO and its short- and long-term evolution as a QPO triplet.

\subsection{Basic Scenario}\label{sec_sce}
Once the long-term evolution is attributed to the {LT} effect, it naturally explains the QPO lag reversal along with other observed long-term evolutionary features.
As shown in panel (a) of Figure~\ref{fig4}, we follow \citet{2009MNRAS.397L.101I} but adopt a truncated {disk} and two-component corona geometry, rather than their hot inner {accretion flow. The entire} corona is assumed to precess around the black hole spin axis in such a way that the misalignment angle between the corona and the spin axis, $\beta$, remains constant.
Since the {disk} is stationary and misaligned with the spin axis by the same angle $\beta$, the angle between the rotation axes of the {disk} and the corona varies from $0$ to $2\beta$ over each precession cycle \citep{2019NewAR..8501524I}.

{We assume that} the observer is located at the top-right corner of the schematics in panels (b)–(h) of Figure~\ref{fig4}.
When the two coronae are tilted away from the line of sight due to the precession effect (panel b), the QPO photons would first interact with the warm corona and then with the hot corona, resulting in a QPO hard lag.
The light travel path difference is approximately $30R_g$, corresponding to a time lag of about 300 seconds at the speed of light, comparable in order of magnitude to the {maximum observed lag of 500 s (see interval A in the top panel of Figure~\ref{fig2})}.
The discrepancy between this lag and the observed 500 s may result from gravitational redshift or from the hot corona being elevated above the disk, resembling a suspended lamp.
When the corona lies flat along the disk plane {(see panel c)}, photons emitted from the QPO region are less likely to interact with the warm corona, rendering the QPO lag undetectable and substantially reducing the QPO amplitude in the soft X-ray band {(see interval B in Figure~\ref{fig2})}.
Meanwhile, the projected area of the hot corona facing the observer increases, resulting in enhanced hard X-ray flux.
Additionally, radiation pressure may push the QPO-emitting region outward to larger radii, leading to a decrease in the QPO frequency.
As the corona tilts toward the observer {(see panel d)}, photons from the QPO-emitting region can no longer illuminate the near-side warm corona. Instead, they interact first with the hot corona, producing the hard X-ray QPO {(see interval C in Figure~\ref{fig2})}.
The corresponding light travel path difference is approximately 40 $R_g$, translating to a delay of over 400 seconds, broadly consistent with the observed {500-s soft} lag.

The origin and evolution of the lag between the hard QPO amplitude and the smoothed soft-band flux (CCF lag; \citealt{2025ApJ...983...13X}) can be broadly explained by extending the current model to include {quasiperiodic ultra-fast outflows (QPOuts)}.
The presence of the ultra-fast outflow {(UFO)} has been suggested by spectral analyses \citep{2024A&A...687A.179X} and is thought to originate from the disk surface, accelerated by radiation pressure \citep{2025ApJ...979..101T}.
We assume that an outburst, modulated by periastron precession, triggers matter to leave the surface of the disk or corona, thereby launching the outflow.
The QPO weakens as the outflow obscures the seed photons from the QPO region (panels e and g). Subsequently, the outflow accelerates and eventually crosses our line of sight, primarily absorbing soft X-ray photons and producing a significant time lag consistent with the observed CCF lag (panels f and h).
When the coronae precess toward us (panels g and h), the angle between the acceleration direction and {our} line of sight is smaller.  
As a result, the observed signal experiences a longer CCF lag compared to when the coronae precess away from us (panels e and f).
Future observations with improved cadence and temporal coverage will be essential to test and refine this model.

\section{Spectral Evidence}\label{sec_spec}

\begin{figure}[t]
    \centering
    \includegraphics[scale=0.45]{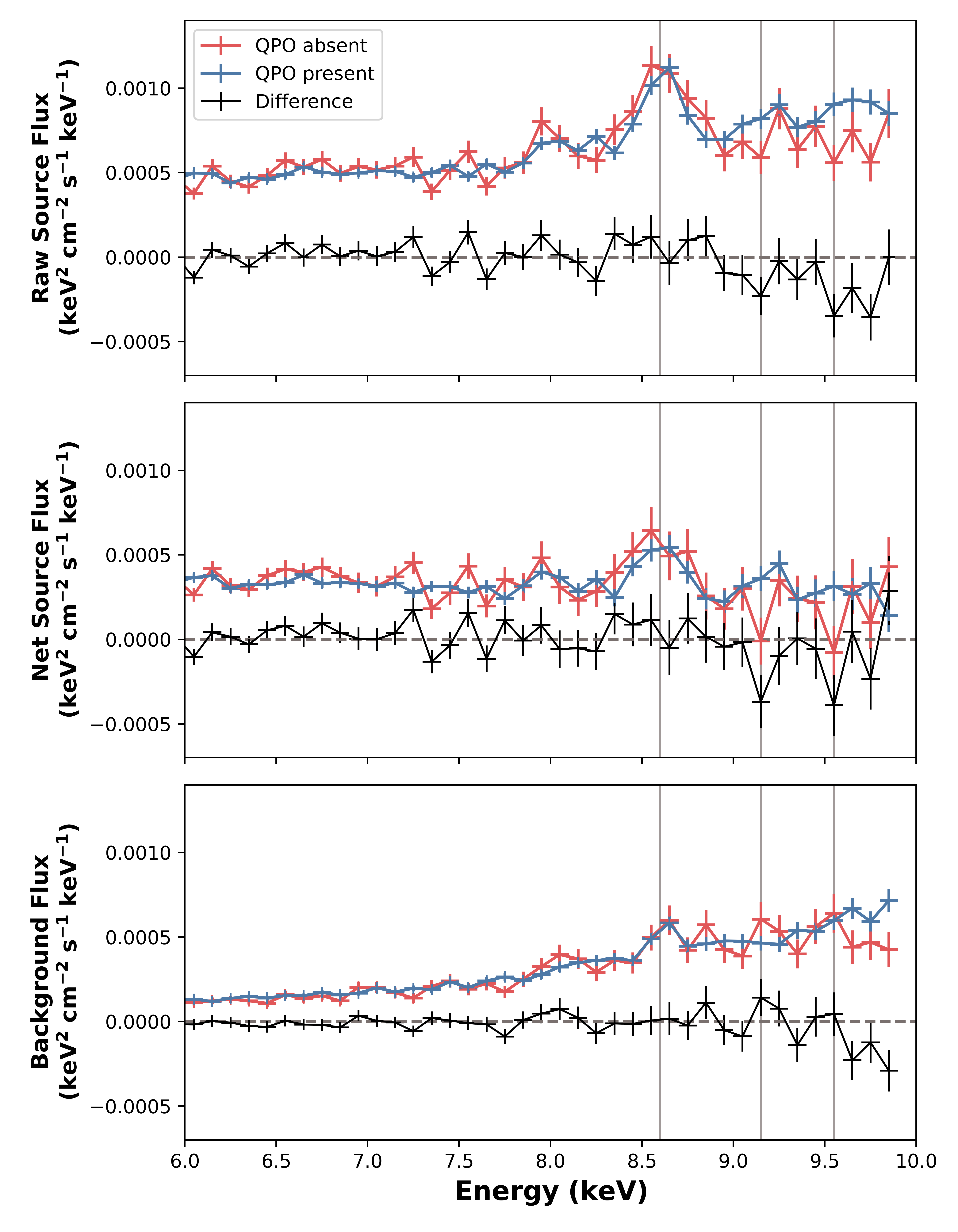}
    \caption{Stacked energy spectra when the QPO is present (blue) and absent (red), and their difference (black; {defined as the QPO-absent spectra minus the QPO-present spectra}), shown for the raw source spectra (top panel), background-{subtracted} source spectra (middle panel), and background spectra (bottom panel). Vertical gray lines at 8.6 keV, 9.15 keV, and 9.55 keV indicate possible features corresponding to a blueshifted $\rm Fe\ K\alpha$ (6.40 keV) emission line, and $\rm Fe\ XXV\ He\alpha$ (6.72 keV) and $\rm Fe\ XXVI\ Ly\alpha$ (7.00 keV) absorption lines, respectively, sharing the same blueshift velocity of $\sim 0.3c$. 
    The dashed horizontal line marks the zero flux.
    }
    \label{fig5}
\end{figure}

\citet{2025ApJ...979..101T} reported a {UFO} in RE~J1034+396 with a line-of-sight velocity of $\sim 0.3c$, evidenced by a significant emission line at $\sim 8.5\ \rm keV$, interpreted as a blueshifted $\rm Fe\ K\alpha$ emission line from outflowing gas. 
A photoionization model scan also identified a similar UFO in this source \citep{2024A&A...687A.179X}.

Based on the wavelet analysis, the intervals during which the hard QPO is present and absent have been identified \citep{2025ApJ...983...13X}. 
For each of the 12 observations in \citet{2025ApJ...983...13X}, the spectra are extracted from the EPIC-pn data corresponding to the two different time intervals.
The source and background spectra are obtained using the \textsc{especget} task, with extraction regions identical to those used for the light curves \citep{2025ApJ...983...13X}. These spectra are then grouped using \textsc{grppha} with a fixed bin width of 20 channels. 
Following the approach of \citet{2024A&A...687A.179X}, we perform spectral combination using the \textsc{epicspeccombine} task. 
However, instead of combining the data based on flux levels, we adopt the QPO-present and QPO-absent intervals identified by \citet{2025ApJ...983...13X}.

It has been noted that the background level is comparable to the source in the high-energy range ($>9\ \rm keV$) \citep{2025ApJ...979..101T}. To constrain the influence of the background, we consider the raw source, background-{subtracted} source, and background spectra, ensuring both convenience and reliability. A power-law model is then used to fit the stacked spectral continuum, and the unfolded spectra are shown in Figure~\ref{fig5}. 
To compare the differences between intervals when the hard QPO is present and absent, we subtract the corresponding spectra, revealing notable differences in the absorption features around $\sim 9.15\ \rm keV$ and $\sim 9.55\ \rm keV$.
Considering a blueshift of $0.3c$, these two absorption features correspond exactly to the $\rm Fe\ XXV\ He\alpha$ and $\rm Fe\ XXVI\ Ly\alpha$ lines, which, together with the $\rm Fe\ K\alpha$ line at 8.6 keV, form a set of outflow features similar to those observed in the spectra of NGC 4051 ({see Figure 6} of \citealt{2015ARA&A..53..115K}), but at higher blueshifted velocities.
{Note that the redshift of RE~J1034+396 is relatively small ($z = 0.043$; \citealt{2016A&A...594A.102C}) and can be neglected in our analysis.
}
The detection of absorption lines during the absence of the QPO aligns with our model ({see} Section~\ref{sec_model}), as absorption features arise only when the outflow lies along our line of sight.
This suggests {a QPOut} that is likely associated with the short-term evolution of the QPO, potentially obscuring both the QPO photons originating from the disk and soft X-ray photons, which results in {reduced} soft flux. Such a mechanism naturally explains the unexpected association between the QPO and the soft X-ray flux, as reported by \citet{2025ApJ...983...13X}.


\section{Discussion}\label{sec_dis}
\subsection{Mass and Spin Measurements}
The central black hole mass of RE~J1034+396 has been estimated using a variety of techniques, resulting in a broad range of values. 
Most measurements place the mass between $10^6$ and $10^7\ M_\odot$ \citep{2016A&A...594A.102C}.
The estimation of $M_{\rm BH} \sim 1.7 \times 10^6 M_\odot$ from the RPM assumption is closely consistent with that measured by stellar velocity dispersion ($\sim 2 \times 10^6 M_\odot$; \citealt{2009ASPC..408..303B,2010MNRAS.401..507B}) among these measurements, marked as grey area in Figure~\ref{fig3}.

Though luminous AGNs are typically associated with rapidly spinning black holes \citep{2014SSRv..183..277R}, NLS1s show a more uniform spin distribution with systematically lower values than Sy1 AGNs \citep{2023Univ....9..175P}.  
Current constraints on the spin of RE~J1034+396 remain inconclusive, although broadband modeling by \citet{2016A&A...594A.102C} favors a low spin. This appears contradictory with the {LT} mechanism requiring Kerr spacetime while our spin estimate approaches zero.  
Notably, tilted structures can form even when $a=0$ \citep{2018MNRAS.474L..81L}. The key question is whether {LT} precession can dominate irradiation variability in such cases.  
Current evidence for low spin remains tentative as it primarily depends on the assumed long-term evolution period. 
Further observations are required to constrain the spin parameter reliably.


\subsection{Origin of QPO}

Previous works have found that the frequency \citep{2024ApJ...961L..32X} and the intervals of the peaks of the QPO patterns \citep{2025ApJ...983...13X} are nearly identical in both the soft and hard X-ray bands, suggesting a common origin.
Two scenarios can be considered. 
The first posits that both soft and hard QPOs are manifestations of oscillations originating from a common QPO region. 
The second proposes that the QPO originates within one coronal component and propagates to another.
However, observations of QPO lag reversals between the soft and hard bands \citep{2024ApJ...961L..32X} challenge the second scenario.
If the QPO were generated within one part of the corona, the source region would need to swing between components to produce the observed reversals. 
This interpretation is difficult to reconcile with a physically plausible mechanism.
In contrast, assuming a disk origin for the QPO is consistent with the {RPM} and naturally accounts for several observed features of the QPO signal.

{
For simplicity, we describe the QPO region as a stable area located close to the observer in Section~\ref{sec_sce} (see Figure~\ref{fig4}). 
However, the asymmetric geometry is not strictly necessary in our model. 
If the QPO region were located on the opposite side of the disk (i.e., being away from the observer), the resulting geometry would not produce an observable QPO lag for any inclination of the corona, since the lengths of light travel paths would be nearly identical whether being reflected by the hot or warm corona.
Therefore, whether the QPO region forms an annular structure or a point-like source co-rotating with the disk, the time-integrated effect can still produce a QPO lag similar to that observed when the region is close to the observer, albeit with some perturbations. Nevertheless, a rotating QPO region (possibly with its own characteristic timescale) may be a more plausible scenario, naturally accounting for the short-term instability of the QPO lag seen in the wavelet analysis, as shown in the lower panels of Figure 1 in \citet{2025ApJ...983...13X}.
}

{There is currently no observational evidence confirming such structures, nor can the origin of the QPO be determined based on existing data.}
However, the RPM does not require a specific physical origin for the QPO, as it merely relates the observed QPO frequencies to the fundamental orbital frequencies in the relativistic regime.
The physical origin of QPOs remains uncertain even in BHBs, despite the availability of far more extensive observational data compared to AGNs. 
Numerical simulations are needed to better constrain the origin of the QPO triplet and to provide predictions.

\section{Summary}\label{sec_sum}
Previous studies have reported the presence of a QPO and its temporal evolution in RE~J1034+396, highlighting the need for a model that can explain such complex behavior. 
In this work, we propose an exploratory scenario based on the RPM, which accounts for the observed {long-term and short-term} evolution and allows independent estimation of the black hole mass and spin from the temporal features.
The main results are summarized as follows:

\begin{enumerate}
    \item We hypothesize that the QPO undergoes a long-term cyclic evolution, with a period of 92.2 days as inferred from fitting the QPO lags to a sinusoidal model. 
    We propose that the previously reported QPO ($2.65 \times 10^{-4}\ \rm Hz$), along with its short-term ($5.25 \times 10^{-5}\ \rm Hz$) and long-term ($1.26 \times 10^{-7}\ \rm Hz$) variations, may constitute a QPO triplet, potentially the first indication of such a structure in an AGN.
    
    \item By applying the RPM to this triplet, we independently constrain the black hole mass to $1.7^{+0.9}_{-0.8} \times 10^6\ M_\odot$ and the dimensionless spin to $0.017^{+0.028}_{-0.012}$. 
    The inferred mass is consistent with estimates obtained through other independent methods, particularly that based on stellar velocity dispersion, which suggests a mass of approximately $2 \times 10^6\ M_\odot$.
    
    \item We propose a schematic model based on the RPM to explain the strong amplitude modulation of the hard-band X-ray QPO with the soft-band flux, along with the long-term evolution of key observational features such as {the variation} of the CCF lag, the QPO time lag {(including its reversal)}, the QPO frequency, and the detection fraction of the soft QPO.

    \item Spectral features provide several indications supporting this scenario {possibly with QPOuts}, including three emission and absorption lines sharing the same blueshift, consistent with a strong UFO at a velocity of $\sim 0.3c$.
    
\end{enumerate}

\begin{acknowledgments}
{The authors thank great insights from Jianmin Wang, Yefei Yuan, Mouyuan Sun, Shifu Zhu and Yijun Wang.}
This work is based on observations obtained with XMM-Newton, an ESA science mission with instruments and contributions directly funded by ESA Member States and NASA.
R.S.X., Y.Q.X., J.L.W., \& H.L. acknowledge support from the Strategic Priority Research Program of the Chinese Academy of Sciences (grant NO. XDB0550300), the NSFC grants (12025303 and 12393814), and the National Key R\&D Program of China (2023YFA1608100 and 2022YFF0503401).
\end{acknowledgments}

\bibliography{sample631}

\begin{thebibliography}{}
\expandafter\ifx\csname natexlab\endcsname\relax\def\natexlab#1{#1}\fi
\providecommand{\url}[1]{\href{#1}{#1}}
\providecommand{\dodoi}[1]{doi:~\href{http://doi.org/#1}{\nolinkurl{#1}}}
\providecommand{\doeprint}[1]{\href{http://ascl.net/#1}{\nolinkurl{http://ascl.net/#1}}}
\providecommand{\doarXiv}[1]{\href{https://arxiv.org/abs/#1}{\nolinkurl{https://arxiv.org/abs/#1}}}

\bibitem[{{Bian}(2009)}]{2009ASPC..408..303B}
{Bian}, W. 2009, in Astronomical Society of the Pacific Conference Series, Vol.
  408, The Starburst-AGN Connection, ed. W.~{Wang}, Z.~{Yang}, Z.~{Luo}, \&
  Z.~{Chen}, 303

\bibitem[{{Bian} \& {Huang}(2010)}]{2010MNRAS.401..507B}
{Bian}, W.-H., \& {Huang}, K. 2010, \mnras, 401, 507,
  \dodoi{10.1111/j.1365-2966.2009.15662.x}

\bibitem[{{Czerny} {et~al.}(2010){Czerny}, {Lachowicz}, {Dov{\v{c}}iak},
  {Karas}, {Pech{\'a}{\v{c}}ek}, \& {Das}}]{2010A&A...524A..26C}
{Czerny}, B., {Lachowicz}, P., {Dov{\v{c}}iak}, M., {et~al.} 2010, \aap, 524,
  A26, \dodoi{10.1051/0004-6361/200913724}

\bibitem[{{Czerny} {et~al.}(2016){Czerny}, {You}, {Kurcz},
  {{\'S}redzi{\'n}ska}, {Hryniewicz}, {Niko{\l}ajuk}, {Krupa}, {Wang}, {Hu}, \&
  {{\.Z}ycki}}]{2016A&A...594A.102C}
{Czerny}, B., {You}, B., {Kurcz}, A., {et~al.} 2016, \aap, 594, A102,
  \dodoi{10.1051/0004-6361/201628103}

\bibitem[{{Das} \& {Czerny}(2011)}]{2011MNRAS.414..627D}
{Das}, T.~K., \& {Czerny}, B. 2011, \mnras, 414, 627,
  \dodoi{10.1111/j.1365-2966.2011.18427.x}

\bibitem[{{Gierli{\'n}ski} {et~al.}(2008){Gierli{\'n}ski}, {Middleton}, {Ward},
  \& {Done}}]{2008Natur.455..369G}
{Gierli{\'n}ski}, M., {Middleton}, M., {Ward}, M., \& {Done}, C. 2008, \nat,
  455, 369, \dodoi{10.1038/nature07277}

\bibitem[{{Ingram} {et~al.}(2009){Ingram}, {Done}, \&
  {Fragile}}]{2009MNRAS.397L.101I}
{Ingram}, A., {Done}, C., \& {Fragile}, P.~C. 2009, \mnras, 397, L101,
  \dodoi{10.1111/j.1745-3933.2009.00693.x}

\bibitem[{{Ingram} \& {Motta}(2019)}]{2019NewAR..8501524I}
{Ingram}, A.~R., \& {Motta}, S.~E. 2019, \nar, 85, 101524,
  \dodoi{10.1016/j.newar.2020.101524}

\bibitem[{{Jin} {et~al.}(2020){Jin}, {Done}, \& {Ward}}]{2020MNRAS.495.3538J}
{Jin}, C., {Done}, C., \& {Ward}, M. 2020, \mnras, 495, 3538,
  \dodoi{10.1093/mnras/staa1356}

\bibitem[{{Jin} {et~al.}(2021){Jin}, {Done}, \& {Ward}}]{2021MNRAS.500.2475J}
---. 2021, \mnras, 500, 2475, \dodoi{10.1093/mnras/staa3386}

\bibitem[{{King} \& {Pounds}(2015)}]{2015ARA&A..53..115K}
{King}, A., \& {Pounds}, K. 2015, \araa, 53, 115,
  \dodoi{10.1146/annurev-astro-082214-122316}

\bibitem[{{Lense} \& {Thirring}(1918)}]{1918PhyZ...19..156L}
{Lense}, J., \& {Thirring}, H. 1918, Physikalische Zeitschrift, 19, 156

\bibitem[{{Liska} {et~al.}(2018){Liska}, {Hesp}, {Tchekhovskoy}, {Ingram}, {van
  der Klis}, \& {Markoff}}]{2018MNRAS.474L..81L}
{Liska}, M., {Hesp}, C., {Tchekhovskoy}, A., {et~al.} 2018, \mnras, 474, L81,
  \dodoi{10.1093/mnrasl/slx174}

\bibitem[{{Liu} {et~al.}(1999){Liu}, {Yuan}, {Meyer}, {Meyer-Hofmeister}, \&
  {Xie}}]{1999ApJ...527L..17L}
{Liu}, B.~F., {Yuan}, W., {Meyer}, F., {Meyer-Hofmeister}, E., \& {Xie}, G.~Z.
  1999, \apjl, 527, L17, \dodoi{10.1086/312383}

\bibitem[{{Maitra} \& {Miller}(2010)}]{2010ApJ...718..551M}
{Maitra}, D., \& {Miller}, J.~M. 2010, \apj, 718, 551,
  \dodoi{10.1088/0004-637X/718/1/551}

\bibitem[{{Middleton} {et~al.}(2011){Middleton}, {Uttley}, \&
  {Done}}]{2011MNRAS.417..250M}
{Middleton}, M., {Uttley}, P., \& {Done}, C. 2011, \mnras, 417, 250,
  \dodoi{10.1111/j.1365-2966.2011.19185.x}

\bibitem[{{Motta} {et~al.}(2014){Motta}, {Belloni}, {Stella},
  {Mu{\~n}oz-Darias}, \& {Fender}}]{2014MNRAS.437.2554M}
{Motta}, S.~E., {Belloni}, T.~M., {Stella}, L., {Mu{\~n}oz-Darias}, T., \&
  {Fender}, R. 2014, \mnras, 437, 2554, \dodoi{10.1093/mnras/stt2068}

\bibitem[{{Piotrovich} {et~al.}(2023){Piotrovich}, {Buliga}, \&
  {Natsvlishvili}}]{2023Univ....9..175P}
{Piotrovich}, M., {Buliga}, S., \& {Natsvlishvili}, T. 2023, Universe, 9, 175,
  \dodoi{10.3390/universe9040175}

\bibitem[{{Reynolds}(2014)}]{2014SSRv..183..277R}
{Reynolds}, C.~S. 2014, \ssr, 183, 277, \dodoi{10.1007/s11214-013-0006-6}

\bibitem[{{Stella} \& {Vietri}(1998)}]{1998ApJ...492L..59S}
{Stella}, L., \& {Vietri}, M. 1998, \apjl, 492, L59, \dodoi{10.1086/311075}

\bibitem[{{Stella} {et~al.}(1999){Stella}, {Vietri}, \&
  {Morsink}}]{1999ApJ...524L..63S}
{Stella}, L., {Vietri}, M., \& {Morsink}, S.~M. 1999, \apjl, 524, L63,
  \dodoi{10.1086/312291}

\bibitem[{{Taylor} {et~al.}(2025){Taylor}, {Wilkins}, \&
  {Allen}}]{2025ApJ...979..101T}
{Taylor}, C.~S., {Wilkins}, D.~R., \& {Allen}, S.~W. 2025, \apj, 979, 101,
  \dodoi{10.3847/1538-4357/ad9f3e}

\bibitem[{{The Manim Community
  Developers}(2025)}]{The_Manim_Community_Developers_Manim_Mathematical_2025}
{The Manim Community Developers}. 2025, {Manim – Mathematical Animation
  Framework}, v0.19.0.
\newblock \url{https://www.manim.community/}

\bibitem[{{Xia} {et~al.}(2024){Xia}, {Liu}, \& {Xue}}]{2024ApJ...961L..32X}
{Xia}, R., {Liu}, H., \& {Xue}, Y. 2024, \apjl, 961, L32,
  \dodoi{10.3847/2041-8213/ad1bf2}

\bibitem[{{Xia} {et~al.}(2025){Xia}, {Liu}, \& {Xue}}]{2025ApJ...983...13X}
---. 2025, \apj, 983, 13, \dodoi{10.3847/1538-4357/adbe6a}

\bibitem[{{Xu} {et~al.}(2024){Xu}, {Pinto}, {Rogantini}, {Barret}, {Bianchi},
  {Guainazzi}, {Ebrero}, {Alston}, {Kara}, \& {Cusumano}}]{2024A&A...687A.179X}
{Xu}, Y., {Pinto}, C., {Rogantini}, D., {et~al.} 2024, \aap, 687, A179,
  \dodoi{10.1051/0004-6361/202349001}

\end{thebibliography}

\end{document}